\documentclass{aa}  

\usepackage[varg]{txfonts}
\usepackage{graphicx}

\usepackage{amsmath}
\usepackage{txfonts}
\usepackage{natbib}
\bibpunct{(}{)}{;}{a}{}{,} 

\usepackage{xcolor}

\begin{document} 

\title{News in the CODE Catalogue}

\titlerunning{News in the CODE Catalogue}

\authorrunning{M.Królikowska \& P.A.Dybczyński}

\author{Małgorzata Królikowska \inst{1}\fnmsep
        \thanks{\email{mkr@cbk.waw.pl}}
         \and
        Piotr A. Dybczyński
          \inst{2}\fnmsep\thanks{\email{dybol@amu.edu.pl}}
        }

   \institute{
   Centrum Badań Kosmicznych Polskiej Akademii Nauk (CBK PAN), Bartycka 18A, 00-716 Warszawa, Poland
       \and
   Astronomical Observatory Institute, Faculty of Physics, Adam Mickiewicz University, Słoneczna 36, PL60-283 Poznań, Poland
}

   \date{}

  \abstract{We describe here updates and new elements of the Catalogue of Cometary Orbits and their Dynamical Evolution (CODE) that, in its original 2020 version, has been introduced by \citet{kroli-dyb:2020}. Currently, the CODE Catalogue offers rich sets of orbital solutions for almost a complete sample of Oort spike comets discovered between 1900 and 2021. We often offer several orbital solutions based on different nongravitational force models or different observational material treatments. An important novelty is that the 'previous' (at previous perihelion or at 120,000 au from the Sun in the past) and 'next' (at next perihelion or 120,000\,au after leaving the planetary zone) orbits are given in two variants. One with the dynamical model restricted only to the full Galactic tide (with all individual stars omitted) and the second one, where all currently known stellar perturbers are also taken into account. Calculations of the previous and next orbits were performed using the up-to-date StePPeD database of potential stellar perturbers.}
   

   \maketitle

\vspace{-0.2cm}
\section{Introduction}\label{introduction}

The Catalogue of Cometary Orbits and their Dynamical Evolution (CODE Catalogue) is mainly dedicated to Oort Cloud comets that is to long-period comets (hereafter LPCs) with an original semi-major axis greater than 10,000\,au in the pure gravitational (GR) model of motion and is restricted to orbital data, not observations. This Catalogue has been introduced by \citet{kroli-dyb:2020}  and made publicly available as the Web version of the Catalogue\footnote{\tt \tiny https://pad2.astro.amu.edu.pl/CODE/}. Therefore, here we only briefly highlight the key points. 

\vspace{0.4cm}
From the beginning, the CODE Catalogue’s novelty is realised in two general ways:
\begin{itemize}
\item by collecting the orbital solution not only for original, osculating, and future orbits\footnote{strictly speaking all three kinds of orbits are osculating, but at very different epochs: when the comet was 250\,au from the Sun before entering the planetary zone, at the epoch close to perihelion passage, and at an epoch equivalent to 250\,au from the Sun when comet leaves the inner part of the Solar System} but also showing orbits at the previous and next perihelion passages. The first three types of orbits can also be found in other cometary databases, such as  the JPL Small-Body Database Browser\footnote{\tiny{\tt https://ssd.jpl.nasa.gov/tools/sbdb\_lookup.html/}} (JPL; only osculating orbits are given), the Minor Planet Center\footnote{\tiny{\tt https://www.minorplanetcenter.net/db\_search/}} (MPC), and the Nakano Notes\footnote{\tiny{\tt http://www.oaa.gr.jp/˜oaacs/nk.htm}}. The last two types of orbits -- to our knowledge -- are only available in the CODE database (for more see Sect.~\ref{sec:previous_next}).
\item by an individual approach to each comet applied to the data treatment and the model of non-gravitational~(NG) acceleration used for the modelling of the particular motion of each comet. Such an individual approach is necessary in the case of comets that disintegrated near their perihelion passage (see Sect.~\ref{sec:comet_disintegration}). In this Catalogue, we generally use three forms of NG~acceleration due to water-ice and CO-ice sublimation. Both aspects are discussed in detail in several of our previous papers; see for example: \citet{kroli-dyb:2012}, \citet{krolikowska:2020}, and \cite{MKLD:2023}. This approach provides several solutions for the majority of comets presented in the CODE Catalogue, in which the preferred solution is highlighted;  what we mean by the preferred orbit in the current edition of our database is described in Sect.~\ref{sec:new_orbits}.  Unfortunately, we do not keep up with comet discoveries due to such a detailed approach to individual objects.
\end{itemize}

\vspace{-0.2cm}
\section{Catalogue contents}\label{sec:CODE-contents}

The CODE Catalogue currently contains orbital solutions for:
\begin{itemize}
\item a complete sample of Oort Cloud comets with q > 3.1\,au from the years 1901-2020.  This set includes 143 objects. Among them there are 6 comets with orbits of 2a or 2b quality class, see \cite{Kroli-Dyb:2018b} for our orbit quality class definitions. Five comets in this sample have a negative $1/a_{\rm ori}$; however, within the 3$\sigma$ uncertainty (three objects -- within 1$\sigma$) these comets may have previously been members of the Solar System.
\item an almost complete sample of Oort spike comets with $q < 3.1$\,au from the same period of 120\,yr. The lack of completeness applies to comets discovered in 2013-2017 ($\sim$20 objects) and four others (C/1958~R1 (Burnham-Slaughter), C/1959~X1 (Mrkos), C/2004~YJ$_{35}$ (LINEAR), and C/2019~N1 (ATLAS)). Currently, 106 comets are in this sample, including 10 with a negative $1/a_{\rm ori}$. Here, 27 comets have orbits of 2a or 2b quality class, and the next 6 comets -- of 3a and 3b quality class. 
\item 54~LPCs with 0.000100\,au$^{-1}$  < $1/a_{\rm ori}$ < 0.000800\,au$^{-1}$. This sample consists almost complete selection of objects with the original semimajor axis close to the boundary of 0.000100\,au$^{-1}$  (say 0.000100\,au$^{-1}$ < $1/a_{\rm ori}$ < 0.000300\,au$^{-1}$).
\item Nine comets that were discovered between 1885--1899 (perihelion distances below 2.0\,au), including six from the list of 19~comets used by Oort to formulate his famous hypothesis. Six of them (four from the original Oort list) are Oort Cloud comets according to the calculation presented in the CODE Catalogue.
\end{itemize}

\begin{figure}
	\centering
	\includegraphics[width=1.04\columnwidth]{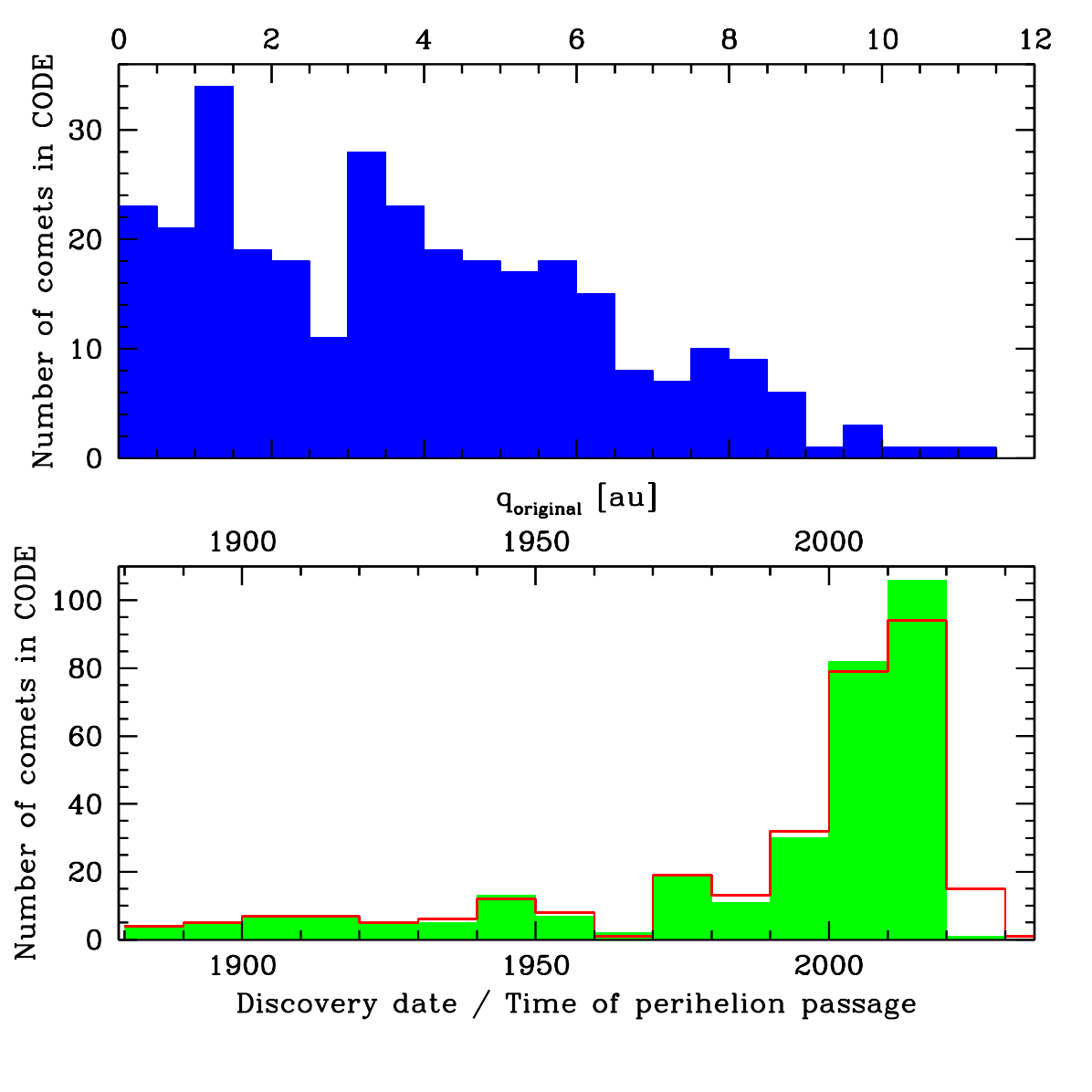}
	\caption{
Characteristics of long-period comets currently in the CODE Catalogue (312 objects). Upper panel: distribution of LPCs as a function of the original perihelion distance. Lower panel: Distribution of LPCs as a function of a discovery date (filled green histogram) and a time of the perihelion passage (red histogram).
        }
	\label{fig:hist1}
\end{figure}

Today in the CODE Catalogue are 766 sets of orbits  (without special user credentials) for 312~LPCs, including 258 Oort Cloud comets. Some characteristics of the entire sample are shown in Fig.~\ref{fig:hist1}.

\vspace{-0.2cm}
\section{New comets in the CODE Catalogue compared to the previous version}\label{sec:new_orbits}

Now, in the CODE Catalogue were added sets of orbits for the complete sample of 29~Oort Cloud comets discovered in  2018, 2019, 2020, and January 2021, and four other LPCs from the same period.  About half of these objects are still observed and some comets are still before their perihelion passage (as of September 2023); see the red histogram in the lower panel of Fig.~\ref{fig:hist1}.  Therefore, for many comets from these years only orbits based on pre-perihelion data are now given in the CODE~Catalogue.  Today, comets from the Oort Cloud are often observed for many years before and after their perihelion passage. Therefore, for many of these comets, it will be necessary to wait a long time before obtaining definitive orbits. This further explains why the CODE Catalogue does not include orbits discovered currently or in very recent years.

\section{A rich palette of new solutions for some comets }\label{sec:newer_orbits}

Recently, \cite{MKLD:2023} selected 32 Oort Cloud comets with large perihelion distances (q> 3.0\,au) observed over the widest ranges of heliocentric distances before and
after perihelion. For each of them, they obtained a series of orbits using at least three basic types of data sets selected from available
positional data (pre- and post-perihelion data and the entire data set), and different forms of NG~acceleration representing water ice or CO
sublimation. Table~2 in \cite{MKLD:2023} lists these comets and all their orbital solutions are now included in the CODE database.

We also added new solutions for several other comets for which data arcs are longer than they were at the moment of orbit determination for the initial CODE Catalogue. This also includes cases of pre-discovery measurements found sometime after the comet's discovery; see for example C/2007~W1 (Boattini), C/2015~R3 (PanSTARRS), and C/2017~K1 (PanSTARRS).

\subsection{The idea of preferred orbits}\label{subsec:preferred_orbits}

In the CODE database, the {\it preferred orbit} is the NG~orbit (or GR~orbit if NG~effects are not satisfactorily detectable in the motion of a given comet) obtained using the longest possible data arc except some untypical cases (for example see Sect.~\ref{sec:comet_disintegration}). Therefore, the preferred orbit best reflects our ability to determine an orbit from all the positional data we have at our disposal. However, to what extent the obtained orbit is well-fitted to the data can be assessed by examining the [O-C] diagrams (Observed - Calculated residuals in right ascension and declination as a function of time) attached to the solutions in the database. There are often visible trends in the data even for the NG~orbit. To understand this, two facts related to non-gravitational effects visible in the motion of Oort Cloud comets should be recalled:

\begin{itemize}
\item The motion of Oort Cloud comets can often be measurably affected by NG~forces even at large heliocentric distances \citep{MKLD:2023}, so the problem is not limited only to comets with small perihelion distances (say below 3\,au).
\item NG~effects are also a source of additional uncertainties resulting from the adopted form of the so-called $g(r)$-like
function. Thus, it is difficult to accurately model NG~accelerations. 
\end{itemize}

\noindent 
Therefore, if the database also gives an orbit based on pre-perihelion data (PRE) or post-perihelion data (POST),the preferred orbit is not always the recommended orbit to study the origin or future of a given comet. In the case of many comets, it was found that the GR~orbit obtained with the data prior to the perihelion is better suited as the starting orbit for the study of previous dynamic evolution than the NG orbit based on the entire data arc.

\subsection{Some details about the set of orbits in CODE Catalogue on the example of comet C/2015~O1 (PanSTARRS)}\label{subsec:c2015o1}

Comet C/2015 O1 (PanSTARRS) (q=3.72\,au) belongs to the group of objects with NG~effects strongly manifested in positional data fitting\footnote{\tiny{\tt ~~https://pad2.astro.amu.edu.pl/CODE/orbit.php?int=2015o1ba \&orb=osculating}}.

The CODE Catalogue presents as many as five orbits based on the entire data arc for C/2015~O1. The first four are purely gravitational orbit (GR, solution 'ba') and three NG~orbits. These NG~orbits, respectively, are based on water ice sublimation ('bn'), CO sublimation ('bc'), and water ice sublimation from the subsolar point ('bs'). All these NG~solutions (based on the symmetric $g(r)$-like function relative to perihelion passage) provide a much better fit of an orbit to the positional data than the GR~orbit; still, however, the fitting to data in 2015 opposition is imperfect. The trends in [O-C] are eliminated only when it is assumed that the maximum $g(r)$ is shifted in time relative to the moment of perihelion passage. Therefore, the asymmetric $g(r)$-like function describing sublimation of water ice from the subsolar point ('bt') is given as the {\it preferred} solution, also because this solution gives the lowest rms within all forms of $g(r)$ tested for this comet; for more details see \cite{MKLD:2023}.

In addition, two orbits based on pre-perihelion data are given ('pa' - GR type solution, 'pc' - NG-CO), an orbit based on part of the pre-perihelion data (heliocentric distance > 5.5\,au, 'pd' - GR), and an orbit based on post-perihelion data arc ('ra' - GR).

Therefore, in this case, it is worth considering three solutions to study the past evolution of this comet: 'bt', 'pc', and 'pd'. It is worth noting that these solutions give slightly different values of $1/a_{\rm ori}$: $40.82\pm 0.56$, $34.85\pm 1.28$, and $33.61\pm 1.83$, in units of $10^{-6}$\,au$^{-1}$, see original orbits for these solutions in the CODE~Catalogue.

For the study of backward evolution, the authors would take the 'pc' or, eventually, the 'pd' orbit as the only GR~orbit that does not give a tendency in [O-C] despite the greatest uncertainties of the orbital elements in this case, and for the study of future evolution -- the 'ra' orbit.

\section{Previous and next orbits -- new approach}\label{sec:previous_next}  

Due to the {\it Gaia} mission, our knowledge on the stars that can pass near the Sun and therefore perturb the motion of LPCs increased enormously. It is now out of the question that the stellar perturbations, together with the simultaneous Galactic action, have to be taken into account when propagating LPC motion far from the Sun, both in the past and in the future. 

But the problem is that the {\it Gaia} catalogues are far from being in their final form. In \cite{First-stars:2020} the first stars that can significantly perturb some of the LPCs were found based on the {\it Gaia} second data release ({\it Gaia}~DR2). A systematic search for such stars performed at that time is described by \cite{rita-pad-magda:2020} and the publicly available Internet database of such stars, StePPeD\footnote{\tt \tiny https://pad2.astro.amu.edu.pl/StePPeD/}, has been announced. But when the {\it Gaia}~EDR3 has been released it was necessary to repeat all this work almost from scratch, see \cite{Dyb-Ber-StePPeD:2022} for details. The reason was that for a large number of stars in question the astrometric results of {\it Gaia}~EDR3 differ significantly from those in DR2. Moreover, the number of known radial velocities increases fourfold in {\it Gaia}~DR3 which makes it necessary to check many more stars for the possibility of a close approach to the Sun. The StePPeD database has been updated several times, the latest version 3.3 was released on 2023~June 28.

In the course of this research, we found that the star HD~7977 might be a strong perturber of most of the LPCs in the past, see \cite{dyb-kroli:2022} for some examples of its influence on particular comets. But according to various {\it Gaia} quality assessments the data for HD~7977 are not conclusive. 

In light of the current uncertainty of many stellar data, both astrometry and radial velocities, we decided that for the CODE catalogue we perform two separate calculations of the 'previous' and 'next' orbits. One with the dynamical model restricted only to the Galactic tide (with all individual stars omitted) and the second one, where all known stellar perturbers' action is also considered. Due to the original interface restrictions, the 'Search' option works only with 'previous' and 'next' orbits obtained from the full dynamical model (stars + Galactic tide).

The second novelty in producing this new CODE version is that for obtaining 'previous' and 'next' orbits we use new and much more precise methods of calculation, described in detail by \cite{Dyb-Breiter:2021}. These methods have been recently updated with the new stellar data, see \cite{Dyb-Ber-StePPeD:2022}.

We also modified the way of presenting the 'previous' and 'next' orbit parameters. In a great majority of cases, the results obtained for the swarm of virtual comets do not resemble the Gaussian distribution. Therefore, we decided to present these results in all cases as three deciles: the tenth, the median, and the ninetieth one.

\section{Comets that do not survive the perihelion passage}\label{sec:comet_disintegration}


As mentioned above, the CODE Catalogue is characterised by an individual approach to orbit determination using positional data. This approach is crucial for comets that disintegrate near their perihelion because the general form of the known NG~models does not adequately reflect the dynamics of the comet. Therefore, for completeness, we decided to describe this aspect using the example of C/2012~S1 (ISON). 

C/2012~S1 was a comet with a very small perihelion distance of 0.0125\,au (about three solar radii). \cite{Sekanina-Kracht:2014} studied the disintegration process of this comet in detail and concluded that the comet fully disintegrated hours before a perihelion passage (expected on 2013 November~28). The first detected minor outburst occurred 16 days before perihelion (at the beginning of November) when the comet was at a distance of 0.7\,au from the Sun. However, an unusual fading of the originally very active comet was reported by observers many weeks before this event.

Thus in the CODE catalogue, two pairs of orbital solutions are offered. The first pair are GR and NG orbits obtained using the entire available data arc, that is over a time interval from  2011 September~30 to 2013 November~11 (a range of heliocentric distance: 9.39\,au -- 1.37\,au). Similarly, the JPL\footnote{The data were retrieved from JPL and MPC orbital databases in 2023~October~17.} presents NG~orbit using the same data arc as above. The MPC gives two NG~orbits (for two different epochs) which are based on different data arcs but both solutions use measurements up to 2013 November~11. All these five solutions are based on data taken while the comet was already in the disintegration process. However, the NG~models of motion used in all three databases do not consider the decay process. 

Therefore in the CODE catalogue the second pair of orbits is based on data taken at large heliocentric distances (more than 3.4\,au from the Sun, these are 'pa' and  'pc' solutions). The first orbit is of a GR type and the second one is an NG orbit obtained using the $g(r)$-like formula describing the CO sublimation. This last orbit was taken as the preferred orbit in the CODE catalogue. 

Similarly, the preferred orbits are obtained for other comets, which we knew were completely or partially disintegrated near the perihelion; see for example C/2012~T5 (Bressi), C/2010~X1 (Elenin), and C/2002 O4 (Hoenig).



\bibliographystyle{aa}   
\bibliography{PAD32} 

\begin{thebibliography}{11}
\expandafter\ifx\csname natexlab\endcsname\relax\def\natexlab#1{#1}\fi

\bibitem[{{Dybczy{\'n}ski} {et~al.}(2022){Dybczy{\'n}ski}, {Berski}, {Tokarek},
  {Podlewska-Gaca}, {Langner}, \& {Bartczak}}]{Dyb-Ber-StePPeD:2022}
{Dybczy{\'n}ski}, P.~A., {Berski}, F., {Tokarek}, J., {et~al.} 2022, \aap, 664,
  A123

\bibitem[{{Dybczy{\'n}ski} \& {Breiter}(2022)}]{Dyb-Breiter:2021}
{Dybczy{\'n}ski}, P.~A. \& {Breiter}, S. 2022, A\&A, 657, A65

\bibitem[{{Dybczy{\'n}ski} \& {Kr{\'o}likowska}(2022)}]{dyb-kroli:2022}
{Dybczy{\'n}ski}, P.~A. \& {Kr{\'o}likowska}, M. 2022, A\&A, 660, A100

\bibitem[{{Kr{\'o}likowska}(2020)}]{krolikowska:2020}
{Kr{\'o}likowska}, M. 2020, \aap, 633, A80

\bibitem[{{Kr{\'o}likowska} \& {Dones}(2023)}]{MKLD:2023}
{Kr{\'o}likowska}, M. \& {Dones}, L. 2023, \aap, 678, A113

\bibitem[{{Kr{\'o}likowska} \& {Dybczy{\'n}ski}(2018)}]{Kroli-Dyb:2018b}
{Kr{\'o}likowska}, M. \& {Dybczy{\'n}ski}, P.~A. 2018, \mnras, 477, 2393

\bibitem[{{Kr{\'o}likowska} \& {Dybczy{\'n}ski}(2020)}]{kroli-dyb:2020}
{Kr{\'o}likowska}, M. \& {Dybczy{\'n}ski}, P.~A. 2020, \aap, 640, A97

\bibitem[{Kr{\'o}likowska {et~al.}(2012)Kr{\'o}likowska, Dybczy{\'n}ski, \&
  Sitarski}]{kroli-dyb:2012}
Kr{\'o}likowska, M., Dybczy{\'n}ski, P.~A., \& Sitarski, G. 2012, A\&A, 544,
  A119

\bibitem[{{Sekanina} \& {Kracht}(2014)}]{Sekanina-Kracht:2014}
{Sekanina}, Z. \& {Kracht}, R. 2014, arXiv e-prints, arXiv:1404.5968

\bibitem[{{Wysocza{\'n}ska} {et~al.}(2020{\natexlab{a}}){Wysocza{\'n}ska},
  {Dybczy{\'n}ski}, \& {Kr{\'o}likowska}}]{First-stars:2020}
{Wysocza{\'n}ska}, R., {Dybczy{\'n}ski}, P.~A., \& {Kr{\'o}likowska}, M.
  2020{\natexlab{a}}, \mnras, 491, 2119

\bibitem[{{Wysocza{\'n}ska} {et~al.}(2020{\natexlab{b}}){Wysocza{\'n}ska},
  {Dybczy{\'n}ski}, \& {Poli{\'n}ska}}]{rita-pad-magda:2020}
{Wysocza{\'n}ska}, R., {Dybczy{\'n}ski}, P.~A., \& {Poli{\'n}ska}, M.
  2020{\natexlab{b}}, \aap, 640, A129

\end{thebibliography}

\end{document}